\documentclass[twocolumn,showpacs,preprintnumbers,amsmath,amssymb,floatfix,prd,superscriptaddress]{revtex4}

\usepackage{graphicx}     
\usepackage{dcolumn}      
\usepackage{bm}           
\usepackage{hyperref}
\usepackage{url}
\usepackage{amsfonts}
\usepackage{mathrsfs}
\usepackage{rsfso}
\usepackage{caption}
\usepackage{subcaption}
\usepackage{comment}
\usepackage{float}
\usepackage{xcolor}

\hypersetup{colorlinks=true, citecolor=blue, linkcolor=blue, urlcolor=blue}

\begin{document}

\title{Exact Gravastar Solution}
\author{Farook Rahaman}
\email[Email: ]{rahaman@associates.iucaa.in}
\affiliation{Department of Mathematics, Jadavpur University, Kolkata 700032, West Bengal, India}

\author{Bikramarka S. Choudhury}
\email[Email: ]{bikramarka@gmail.com}
\affiliation{Department of Mathematics, Jadavpur University, Kolkata 700032, West Bengal, India}

\author{Aritra Sanyal}
\email[Email: ]{aritrasanyal1@gmail.com}
\affiliation{Department of Mathematics, Jadavpur University, Kolkata 700032, West Bengal, India}

\author{Anikul Islam}
\email[Email: ]{anikulislam0025@gmail.com}
\affiliation{Department of Mathematics, Jadavpur University, Kolkata 700032, West Bengal, India}

\author{Bidisha Samanta}
\email[Email: ]{samantabidisha@gmail.com}
\affiliation{Department of Mathematics, Jadavpur University, Kolkata 700032, West Bengal, India}

\begin{abstract}
    Astrophysical black holes come from exact solutions of the Einstein field equations. Therefore, any alternative—such as a gravastar—must meet the same standard of mathematical accuracy and internal consistency. A credible gravastar model cannot rely on approximations or simply patch together different regions. It must provide a single, exact, and self-consistent solution to the Einstein field equations throughout the entire spacetime. 
    We propose an exact solution to the Einstein Field Equations in the context of Gravitationally Vacuum Star, or Gravastar, introduced by  Mazur Mottola \cite{manzur_motolla}. This concept suggests an alternative scenario to the end state of a gravitational collapse indicating the formation of a compact body other than a black hole. Our approach realizes this concept by partitioning the Gravastar into three distinct regions, each defined by exact solutions to the Einstein Field Equations.   We explore the key physical properties of gravastar models, which have been proposed as possible alternatives to classical black holes. This study offers a clear and thorough examination of their theoretical structure and evaluates how physically reasonable gravastars are as compact astrophysical objects. 
\end{abstract}

\maketitle

\section{Introduction}

Einstein's general relativity \cite{GR001} revolutionizes our understanding of gravity by describing it not as a force but as a consequence of the curvature of spacetime caused by matter and energy distribution. The cornerstone of this theory lies in the Einstein field equations, which mathematically relate spacetime curvature to the energy-momentum tensor that encapsulates the matter-energy content. Solutions to these non-linear partial differential equations provide profound insights into the geometric nature of gravity in various astrophysical contexts.

The first exact solution to these equations, known as the Schwarzschild solution \cite{Schwarzschild}, describes a static and spherically symmetric vacuum spacetime surrounding a non-rotating, uncharged mass. This seminal solution reveals the existence of event horizons and singularities—regions where the curvature becomes infinite—thereby laying the theoretical foundation for black holes. The black hole \cite{bh01,bh02} concept emerged as a striking consequence of general relativity, portraying objects so dense that not even light can escape their gravitational grasp.

Astrophysical modeling further substantiates black hole formation as a natural end state for massive stars undergoing gravitational collapse \cite{collapse01,collapse02,collapse03}. The Chandrasekhar limit \cite{chandrasekhar}, approximately 1.44 times the mass of the Sun, provides a critical threshold delineating whether a star will end its life as a white dwarf or collapse further. Stars surpassing this mass limit can no longer be supported by electron degeneracy pressure, triggering collapse processes that may lead to supernovae and the birth of neutron stars or black holes.

Despite the widespread acceptance of black holes as astrophysical entities, alternative schools of thought challenge the inevitability of event horizon formation. One such proposal is the gravastar (gravitational vacuum condensate star) \cite{manzur_motolla}, conceptualized as a quantum phase transition alternative to classical black holes. In the Mazur–Mottola model, the collapsing star avoids forming a central singularity or event horizon by transitioning its interior into a de Sitter spacetime dominated by vacuum energy \cite{desitter}. This interior is enveloped by a thin, ultra-relativistic shell of dense matter, smoothly matched to an external Schwarzschild spacetime. This configuration produces a geodesically complete object without an event horizon, potentially resolving paradoxes such as information loss and singularities intrinsic to traditional black holes. 

However, a key limitation of current gravastar models is the lack of an exact mathematical formulation of the Einstein field equations for the shell region. The shell's highly relativistic and ultra-thin nature poses substantial challenges for a consistent description within general relativity's framework. Addressing this gap, our work proposes an exact solution to the Einstein field equations in the shell region, thereby providing a more concrete and mathematically rigorous description of gravastars. This not only reinforces the theoretical viability of gravastars but also opens avenues for further analytical and observational explorations.

Our paper is organized as follows. In section \ref{sec2}, we provide a brief overview of gravastars. Section \ref{sec3} is the mathematical formulation of our model. We discuss the proper thickness and proper energy of the shell region in section \ref{sec4}. Section \ref{sec5} deals with the junction conditions. The surface redshift is discussed in section \ref{sec6}. Section \ref{sec7} is for the causality condition. Then in section \ref{sec8}, we analyse the deflection angle for the thick shell. Section \ref{sec9} is a comprehensive study of the entropy of the system. Finally, Section \ref{sec10} is the concluding remarks and discussions relating to our study.

\section{An overview of gravastars} \label{sec2}

Since astrophysical black holes arise as exact solutions of the Einstein field equations, any proposed alternative to black holes, i.e. gravastars must satisfy the same standard of mathematical precision and internal consistency. A viable gravastar model, therefore, cannot depend on approximations or loosely joined segments; it must offer an exact and fully self-consistent solution to the Einstein field equations across the whole spacetime.

In the gravastar framework, spacetime is divided into three distinct regions, each characterized by its own specific physical properties. 

\begin{enumerate}
    \item Interior region – typically described by a de Sitter–like spacetime with negative pressure.
    \item Shell region (thin or thick) – where the matter distribution must itself be described by an exact solution of the Einstein field equations, supported either by a physically realistic matter profile or an appropriate exotic equation of state. 
    \item Exterior region – The spacetime outside the gravastar should match an exact vacuum solution of the EFE—typically the Schwarzschild solution for an isolated, non-rotating, spherically symmetric configuration.
    \item Each region must individually satisfy the EFEs exactly, with an appropriate stress–energy tensor.
    \item The junction conditions (Israel–Darmois conditions) must be satisfied at the shell–exterior interface to ensure that the overall spacetime remains smooth and physically consistent.
    \item The resulting configuration must avoid horizons and singularities, while still reproducing the black hole–like exterior geometry required for observational viability.
\end{enumerate}

The very first description of gravitationally vacuum condensate stars was by Mazur and Mottola in their paper \cite{manzur_motolla}. Later, gravastars have been the subject of considerable theoretical investigation within the context of alternative gravity models and compact objects. Das et al.\ have studied gravastar configurations in the framework of $f(R,\mathscr{T})$ gravity \cite{grava01}. Study of gravastars in Rastall gravity has been done by Shounak Ghosh et al.\ in \cite{grava07}. Ghosh et al.\ have investigated charged gravastars in higher-dimensional spacetimes, expanding the theoretical landscape of these objects and examining the influence of higher dimensions on their stability and physical characteristics \cite{grava02}. The formation and observational aspects of gravastar shadows have been analyzed by Sakai, Nobuyuki et al., offering insight into the possible observational signatures that are specific to gravastars \cite{grava03}. Wang, Yu-Tong et al.\ considered the notion of primordial gravastars from an inflationary cosmology perspective, \cite{grava04}. Work by Camilo Posada has further extended gravastar theory to include slowly rotating super-compact objects, thus accounting for astrophysically relevant rotational effects in gravastar models \cite{grava05}. Shin'ichi Nojiri and G.G.L. Nashed have studied a stable gravastar with a large surface redshift in Einstein's gravity with two scalar fields in their work \cite{grava06}.

\section{Mathematical Formalism}  \label{sec3}
We take the spherically symmetric line element as:
\begin{equation}
    ds^2 = -e^{\nu} dt^2 + e^{\lambda} dr^2 + r^2 d\Omega_2^2
\end{equation}

The Einstein Field Equations for an isotropic perfect fluid distribution are given as:
\begin{equation} \label{efe1}
    8 \pi \rho = e^{-\lambda} \left( \frac{\lambda'}{r} - \frac{1}{r^2} \right) + \frac{1}{r^2}
\end{equation}
\begin{equation} \label{efe2}
    8 \pi p = e^{-\lambda} \left( \frac{1}{r^2} + \frac{\nu'}{r} \right) - \frac{1}{r^2}
\end{equation}
{
\begin{equation} \label{efe3}
    8 \pi p = \frac{1}{2} e^{-\lambda} \left[ \frac{{\nu'}^2}{2} + \nu'' - \frac{1}{2} \lambda' \nu' + \frac{1}{r} (  \nu' - \lambda') \right]
\end{equation}}
\noindent

The energy-momentum conservation equation $T^{\mu \nu}{}_{;\nu} = 0$ yields the well-known Tolman–Oppenheimer–Volkoff (TOV) equation:
\begin{equation} \label{tov}
    \frac{dp}{dr} = -\frac{1}{2} (\rho + p) \frac{d\nu}{dr}
\end{equation}

\vspace{1em}

The system is divided into three regions depending on the equation of state (EoS):

\begin{align*}
&\text{(i)\,\, de-Sitter Core: } &&p = -\rho, \quad 0 \le r < r_1,\\
&\text{(ii)\,\, Thick Shell: } &&p = -\tfrac{1}{5}\rho, \quad r_1 < r < r_2,\\
&\text{(iii)\,\, Schwarzschild Exterior: } &&p = \rho = 0, \quad r_3 \le r.
\end{align*}

\subsection{Interior Space-time}

The matter distribution in the interior region is characterized by the equation of state:
\begin{equation}
    p = -\rho
\end{equation}

Using Eqs.~(\ref{efe1}) and (\ref{efe2}), we obtain:
\[ 
    \lambda' = -\nu'
\]
Hence, we find the ﬁnal expression for metric potential  as,
\begin{align}&e^{-\lambda} =  1-H^2 r^2 \\
&e^\nu = C_1( 1-H^2 r^2)
\end{align}
where $H$ and $C_1$  are  constants.

\subsection{Thick Shell Region}

In this region, the equation of state is taken as:
\begin{equation}
    p = -\frac{1}{5} \rho
\end{equation}

Substituting this relation into the field equations leads to the following set of solutions:\cite{rahaman2008static}

\begin{equation} \label{001}
    e^{\nu} = \frac{K}{r}
\end{equation}
\begin{equation} \label{002}
    e^{\lambda} = \frac{1}{\frac{C}{r} - 4}
\end{equation}
\begin{equation}
    p = -\frac{1}{8 \pi r^2}\label{003}
\end{equation}
\begin{equation}
    \rho = \frac{5}{8 \pi r^2}\label{004}
\end{equation}
where $K$ and $C$ are arbitrary constants.

From Eqs.~(\ref{001}) and (\ref{002}), since both left-hand sides must be positive, the constants and radial range are restricted as:
\begin{equation}
    K > 0, \quad C > 0
\end{equation}
\begin{equation}
    \frac{C}{r} - 4 > 0 \implies r < \frac{C}{4}
\end{equation}

If $r = R$ denotes the radius of the core region, then the radial coordinate of the thick shell region must satisfy:
\begin{equation} \label{restriction01}
    R < r < \frac{C}{4}
\end{equation}
In this work, we fix the outer boundary of the shell region to the limiting value at \( C/4 \). This choice allows greater flexibility in the core-to-shell boundary \( R \). Note that one could instead take the shell's outer boundary as some \( \tilde{R} \) satisfying \( R < r < \tilde{R} < C/4 \). To adapt the results of this paper for such an \( \tilde{R} \), simply replace \( C \) with \( 4\tilde{R} \) throughout the remaining sections.

\subsection{Exterior Space-time}

The exterior is a vacuum solution of the field equations, characterized by:
\begin{equation}
    p = \rho = 0
\end{equation}
Hence, the exterior geometry is described by the Schwarzschild metric:
\begin{equation}
    e^{\nu} = 1 - \frac{2M}{r}, \qquad e^{\lambda} = \frac{1}{1 - \frac{2M}{r}}
\end{equation}
Here, $M$ represents the total mass of the object as a whole.

\section{Proper Thickness and Energy}  \label{sec4}

In this gravastar model, the shell separating the interior de Sitter region and the exterior Schwarzschild region plays a crucial role in determining the structural and energetic properties of the configuration. Two key physical quantities that characterize the shell are its \textit{proper thickness} and \textit{proper energy}. These quantities are computed using the spatial metric of the intermediate region.

\subsection{Proper Thickness}

The \textit{proper thickness} of the shell, denoted by $l$, represents the physical radial distance measured across the shell as experienced by a local observer. It is defined through the metric component $g_{rr}$ as
\begin{equation}
    l = \int_{R}^{C/4} \sqrt{g_{rr}} \, dr
    \label{eq:proper-thickness}
\end{equation}
where $R$ and $C/4$ correspond to the inner and outer coordinate boundaries of the shell, respectively. The metric component $g_{rr}$ is determined by the spacetime geometry in the shell region given by \eqref{002}. Therefore, the expression becomes:
\begin{align}
    l &= \int_{R}^{C/4} \frac{1}{\sqrt{\frac{C}{r}-4}} \,\,dr \nonumber \\
    &= \left[ -\frac{\sqrt{C} \sqrt{r} \sqrt{\frac{C - 4r}{C}}}{4} + \frac{C}{8} \arcsin \left( \frac{2 \sqrt{r}}{\sqrt{C}} \right) \right]_{R}^{\frac{C}{4}} \nonumber \\
    &= \frac{  \sqrt{R(C-4R)}}{4} - \frac{C}{8} \arcsin\left(\frac{2\sqrt{R}}{\sqrt{C}}\right) + \frac{\pi C}{16}
\end{align}

Physically, the integral in Eq.~(\ref{eq:proper-thickness}) accounts for the curvature-induced deformation of radial coordinates, ensuring that $l$ measures the true radial distance between the two surfaces rather than the coordinate separation. For an ultra-thin shell (i.e., when $l \to 0$), the configuration approximates the idealized surface layer of the original Mazur–Mottola gravastar model.

\subsection{Proper Energy}

The \textit{proper energy} contained in the shell, denoted by $E$, measures the total energy stored in the matter distribution between $r = R$ and $r = C/4$. It is obtained from the volume integral of the energy density $\rho(r)$ as
\begin{equation}
    E = \int_{R}^{C/4} 4\pi r^2 \rho(r) \, dr
    \label{eq:proper-energy}
\end{equation}

Here, $\rho(r)$ represents the local energy density of the fluid within the shell, and the integration element $4\pi r^2 dr$ corresponds to the proper volume element in spherical symmetry. Therefore, the expression is given as:
{
\begin{align}
    E &= \int_{R}^{C/4} 4\pi r^2 \times \frac{5}{8 \pi r^2} \, dr \nonumber \\
    &= \frac{5}{2} \int_{R}^{C/4}  \, dr = \frac{5}{2} \left(  \frac{C}{4} - R \right)
\end{align}
}

This expression provides an estimate of the total energy stored in the shell matter. The proper energy $E$ is an essential quantity in determining the total mass–energy content of the gravastar and contributes to the gravitational field experienced externally.

\subsection{Physical Interpretation}

The proper thickness $l$ and proper energy $E$ together describe the internal structure of the gravastar shell. A smaller $l$ corresponds to a sharper transition between the inner and outer regions, while a larger $E$ signifies higher energy concentration in the shell. These parameters are particularly significant when assessing the stability and thermodynamic behavior of the configuration. In realistic models, they are constrained by junction conditions at the interfaces and by pressure balance between the interior and exterior spacetimes.

{
\section{Boundary and Junction Conditions}  \label{sec5}

Let $\Sigma$ be a timelike or spacelike hypersurface embedded in a spacetime with metric $g_{\mu\nu}$. We coordinatize $\Sigma$ by intrinsic coordinates $y^a$, so the embedding is given by $x^\mu = x^\mu(y^a)$, where $a,b,\dots = 0,1,2$ (for a 3‑surface in 4‑dimensional spacetime).

The \emph{first fundamental form} is the induced metric $h_{ab}$ on $\Sigma$, defined as
\[
h_{ab} = g_{\mu\nu} \frac{\partial x^\mu}{\partial y^a} \frac{\partial x^\nu}{\partial y^b}.
\]
The line element on the hypersurface is then $\mathrm{d}s^2|_\Sigma = h_{ab}  \mathrm{d}y^a  \mathrm{d}y^b$. The Darmois--Israel condition requires continuity of the first fundamental form:
\[
[h_{ab}] = 0.
\]
where $[X] = X^{+} - X^{-}$ denotes the jump across the surface. This ensures that the intrinsic geometry of $\Sigma$ is well matched.

Let $n^\mu$ be the unit normal to $\Sigma$, normalized so that $n_\mu n^\mu = \varepsilon = \pm 1$ (with $\varepsilon = -1$ for a timelike $\Sigma$, $\varepsilon = +1$ for a spacelike $\Sigma$). The \emph{second fundamental form} (extrinsic curvature tensor) is given by
\[
K_{ab} = e_a{}^\mu e_b{}^\nu \nabla_\mu n_\nu,
\]
where $e_a^\mu$ is the tangent basis vector to the hypersurface $\Sigma$ and the rest of the symbols have their usual meaning.

\subsection{Surface Energy Density and Pressure}

We consider a thin shell [see \cite{junction0001}] located at the junction hypersurface at $r = a(\tau)$ separating two space--times denoted by ``$+$'' (outer region) and ``$-$'' (inner region). The surface stress-energy tensor is
\begin{equation}
S^{a}{}_{b} = \mathrm{diag}(-\sigma,\, \mathcal{P},\, \mathcal{P}),
\end{equation}
where $\sigma$ is the surface energy density and $p$ is the surface pressure.

The Israel junction condition relates the jump in extrinsic curvature to the surface stress-energy tensor $S_{ab}$,
\begin{equation}
S_{ab} = -\frac{1}{8\pi}\left( [K_{ab}] - \delta_{ab}[K] \right),
\end{equation}

For the static spherically symmetric metic the surface energy density is given by
\begin{equation} \label{surface_density}
\sigma = -\frac{1}{4\pi a}
\left[
\sqrt{e^{-\lambda_{+}}(a) + \dot{a}^{2}}
-
\sqrt{e^{-\lambda_{-}}(a) + \dot{a}^{2}}
\right],
\end{equation}

and the surface pressure is

\begin{align}
   \mathcal{P}=& \frac{1}{8
   \pi  a}\Bigg(\frac{a \ddot{a}+\dot{a}^2 \left(\frac{a \lambda
   _+'}{2}+\frac{a \nu _+'}{2}+1\right)+\frac{1}{2} a
   e^{-\lambda _+} \nu _+'+e^{-\lambda _+}}{\sqrt{a'(\tau
   )^2+e^{-\lambda _+}}}\nonumber\\&-\frac{a \ddot{a}+\dot{a}^2
   \left(\frac{a \lambda _-'}{2}+\frac{a \nu
   _-'}{2}+1\right)+\frac{1}{2} a e^{-\lambda _-} \nu
   _-'+e^{-\lambda _-}}{\sqrt{a'(\tau )^2+e^{-\lambda _-}}}\Bigg)
\end{align}

For the static junction ($\dot{a}=0$, $\ddot{a}=0$) at $\tilde{R}$, the surface energy density $\sigma$ and pressure $\mathcal{P}$ are given by the Lanczos equations:

\begin{equation}
    \sigma = - \frac{1}{4\pi a} \left[  \sqrt{e^{- \lambda_{+}}(a)} - \sqrt{e^{- \lambda_{-}}(a)}  \right]
\end{equation}
\begin{align}
    \mathcal{P} = \frac{1}{8\pi a}\Bigg[ \sqrt{e^{- \lambda_{+}}(a)} - \sqrt{e^{- \lambda_{-}}(a)} \nonumber  \\
    + \frac{a}{2} \bigg( \frac{\nu_{+}'(a)}{\sqrt{e^{\lambda_{+}}(a)}} - \frac{\nu_{-}'(a)}{\sqrt{e^{\lambda_{-}}(a)}}  \bigg) \Bigg]
\end{align}
where primes denote $d/da$, $\nu'_\text{+}(a) = \frac{2M}{a^2(1-2M/a)}$, and $\nu'_\text{-}(a) = -K/a^2$.

We also discuss the potential function by the method of Poisson $\&$ Visser as discussed in their work \cite{stability_potential02}. They essentially take the equation \eqref{surface_density} and rearrange the equation in such a manner that it takes the form
\[ \dot{a}^2 + V(a) = 0  \]
This $V(a)$ is the Poisson–Visser effective potential for junction. With simple calculations, we can arrive at the following expression for $V(a)$:

\begin{equation}
\boxed{
\begin{aligned}
V(a)
=&\;
\frac{1}{2}
\left[
e^{-\lambda_{+}}(a)
+
e^{-\lambda_{-}}(a)
\right]
-
\left(2\pi a\,\sigma(a)\right)^2  \\
&\;
-
\frac{
\left[
e^{-\lambda_{+}}(a)
-
e^{-\lambda_{-}}(a)
\right]^2
}{
64\pi^2 a^2 \sigma^2(a)
}.
\end{aligned}
}
\label{potential}
\end{equation}

The continuity of the first fundamental form can be readily seen from the metric continuity discussed below. Also, the non-vanishing surface stress-energy tensor dictates the jump discontinuity in the second fundamental form. This validates our junctions to be mathematically viable. In the following junction analysis, we assume the junctions are static i.e. $\dot{a},\ddot{a}=0$.

\subsection{Core-Shell Junction at $r = R$}

At the inner boundary $r = R$:
\begin{align}
    C_1(1 - H^2 R^2) &= \frac{K}{R} \\
    (1 - H^2 R^2) &= C/R - 4   \\
    \implies K/C_1 = C - 4R &\implies\boxed{ R =\frac{C-K/C_1}{4}}
\end{align}
This gives us a measure on the core region radius for the formation of a stable gravastar.

For this junction, the non-zero components surface stress-energy tensor is given by:
\begin{equation}
    \sigma = -\frac{1}{4\pi a}\left[\sqrt{\left(\frac{C}{a}-4\right)} - \sqrt{1-a^2H^2}\right]
\end{equation}
\begin{align}
    \mathcal{P} = \frac{1}{8\pi a}\Bigg[\sqrt{\frac{C}{a}-4}-\sqrt{1-a^2H^2}\Bigg]   \nonumber   \\  
    -\frac{1}{16\pi}\Bigg[  \frac{1}{a} \sqrt{\left(\frac{C}{a} - 4\right)} -\frac{2aH^2}{\sqrt{1-H^2a^2}}    \Bigg]  
\end{align}

The Poisson–Visser effective potential given by \eqref{potential} for the core-shell junction is calculated and plotted in Fig.~\ref{fig:int_shell}.

\begin{figure}
    \centering
    \includegraphics[width=0.9\linewidth]{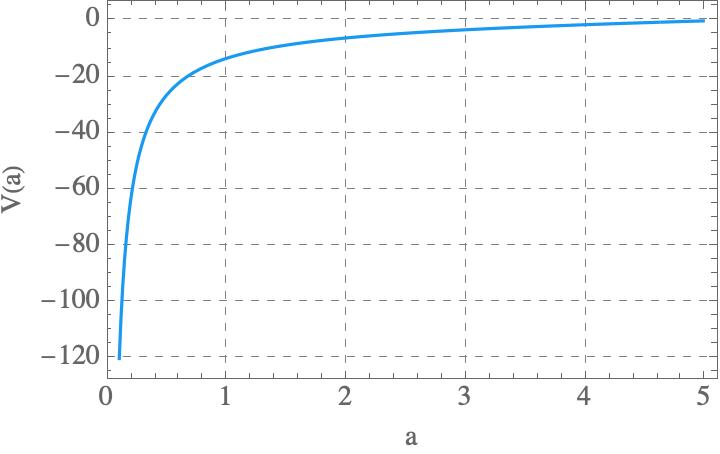}
    \caption{Plot of $V(a)$ vs $a$ for the interior and shell region junction given by the equation \eqref{potential} considering the parameter values of $M,K,C_1=1;\,\,$ $ C=20\,;\,\,H=0.2$. This allows the shell to extend upto radius 5 units and the core-shell junction to be anywhere between 0 to 5 units.}
    \label{fig:int_shell}
\end{figure}

\subsection{Shell-Exterior Junction at $r = \tilde{R}$}

\textbf{$g_{tt}$ continuity:} Match the shell region's $e^{\nu} = K/r$ with the exterior Schwarzschild metric:
\begin{equation}
    \frac{K}{\tilde{R}} = 1 - \frac{2M}{\tilde{R}} \implies K = \tilde{R} \left(1 - \frac{2M}{\tilde{R}}\right)
    \label{eq:nu-matching}
\end{equation}

However, $K > 0$ is required for positive $g_{tt}$. This indicates we need $\tilde{R}>2M$, i.e. the outer boundary radius is greater than the schwarzschild radius.

For this junction:
\begin{align}
    \sigma(a) = - \frac{\sqrt{1-\frac{2 M}{a}}-\sqrt{\frac{C}{a}-4}}{4 \pi  a}
\end{align}
\begin{align}
   \mathcal{P}= \frac{1}{8 \pi a}\left(\frac{1-\frac{M}{a}}{\sqrt{1-\frac{2 M}{a}}}-\frac{1}{2} \sqrt{\frac{C}{a}-4}\right)
\end{align}

The Poisson–Visser effective potential given by \eqref{potential} for the shell-exterior junction is calculated and plotted in Fig.~\ref{fig:shell-ext}.

\begin{figure}
    \centering
    \includegraphics[width=0.9\linewidth]{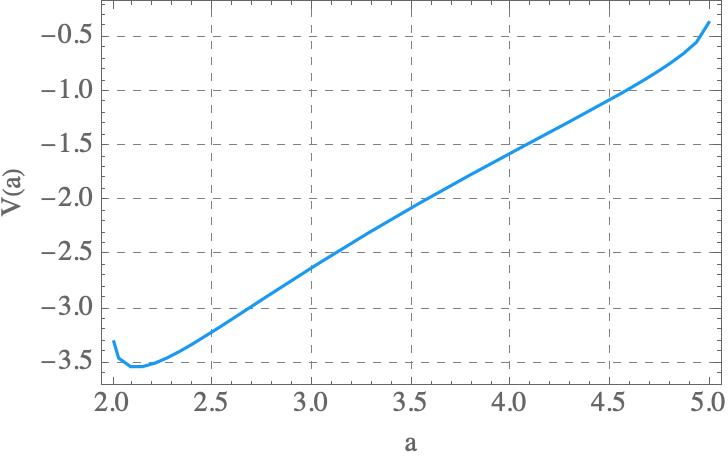}
    \caption{Plot of Poission-Visser potential $V(a)$ vs $a$ for the shell-exterior junction region given by the equation \eqref{potential}. Here we take the parametric values as $M,K,C_1=1;\,\,$ $ C=20\,;\,\,H=0.2$. This provides us with a Schwarzschild radius of 2 units. The $C$-value implies the shell can extend upto 5 units.}
    \label{fig:shell-ext}
\end{figure}

}

\section{Surface Redshift}  \label{sec6}

The investigation of the surface redshift associated with a gravastar provides crucial information regarding its stability and potential observational signatures. The surface gravitational redshift is defined as
\begin{equation}
    Z_s = \frac{\lambda_{0}- \lambda_{e}}{\lambda_{e}} = \frac{\lambda_{0}}{\lambda_{e}}-1
\end{equation}
where $\lambda_{0}$ and $\lambda_{e}$ denote the emitted and observed wavelengths, respectively.
Buchdahl \cite{Buchdahl1959} established that, for a static and isotropic perfect fluid sphere, the surface redshift cannot exceed the value $Z_s$ = 2. For an anisotropic matter distribution, this upper bound may increase to approximately 3.84 \cite{Ivanov2002}. Barraco and Hamity \cite{Barraco2002} demonstrated that in the absence of a cosmological constant, the upper limit $Z_s \le2$ remains valid for isotropic configurations.
The equation for surface redshift for the adopted spacetime metric is given as:
\begin{equation}
    Z_s = -1 + \frac{1}{\sqrt{|g_{tt}|}} = -1 + \sqrt{\frac{r}{K}}
\end{equation}

Two immediate consequences follow from the definition \(Z_s(r,K) = -1 + \sqrt{r/K}\):
\begin{enumerate}
\item Non-blueshift: $Z_s \geq 0 \Leftrightarrow r \geq K$. At $r=K$, $Z_s=0$; for $r<K$, $Z_s<0$.
\item Buchdahl bound: $Z_s < 2 \Leftrightarrow \sqrt{r/K} < 3 \Leftrightarrow r < 9K$.
\end{enumerate}

Thus, if one demands both \(Z_s \ge 0\) and \(Z_s < 2\) at a given radius, the admissible \(r\)-interval at fixed \(K\) is
\[
\boxed{ \quad K \;\le\; r \;<\; 9K \quad }
\]
Equivalently, at fixed \(r\) the admissible \(K\)-interval is
\[
\boxed{ \quad \frac{r}{9} \;<\; K \;\le\; r \quad }
\]
(the left inequality enforces \(Z_s<2\); the right one avoids blueshift).

\begin{figure}[H]
  \centering
  \includegraphics[width=\linewidth]{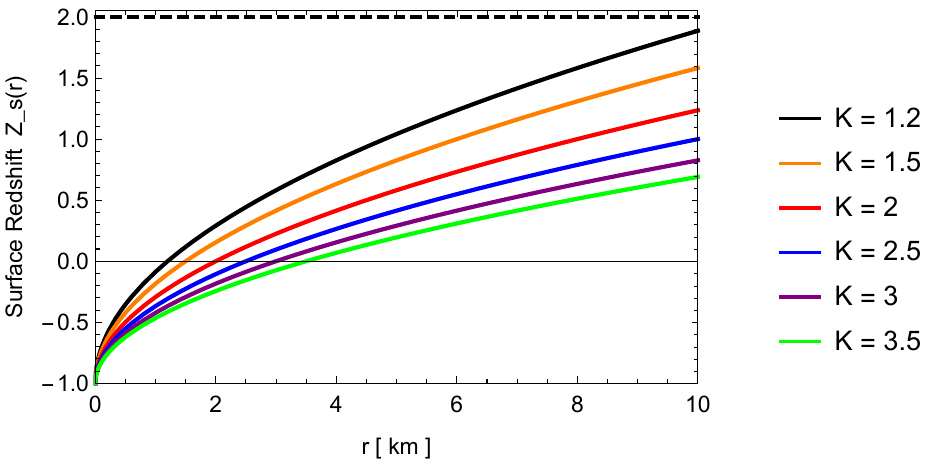}
  \caption{Surface redshift \(Z_s(r)\) for several \(K\). The horizontal line at \(Z_s=2\) indicates the Buchdahl threshold; the segment \(r<9K\) on each curve satisfies \(Z_s<2\).}
  \label{fig:ZsVsR}
\end{figure}

\begin{figure}[H]
  \centering
  \includegraphics[width=0.85\linewidth]{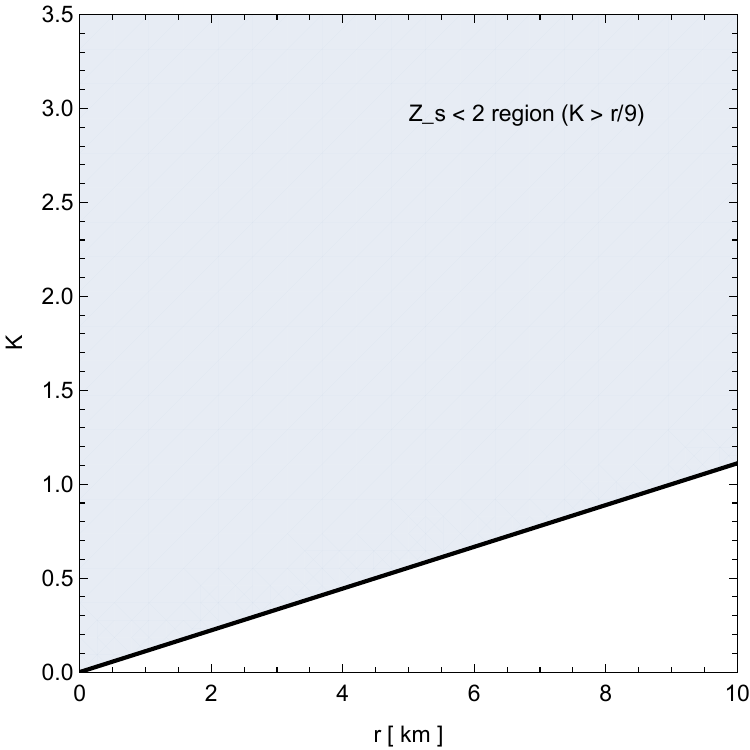}
  \caption{Threshold \(K=r/9\) in the \((r,K)\) plane. The region above the line satisfies \(Z_s<2\). Imposing \(Z_s\ge 0\) adds the upper bound \(K \le r\), yielding \(r/9<K\le r\).}
  \label{fig:KThreshold}
\end{figure}

Fig. \ref{fig:ZsVsR} shows \(Z_s(r)\) for representative \(K\) values; the horizontal line at \(Z_s=2\) marks the threshold, and the curve segments with \(Z_s<2\) (i.e.\ \(r<9K\)) can be emphasised. Fig. \ref{fig:KThreshold} plots the threshold \(K=r/9\); the admissible region for \(Z_s<2\) lies above this line. If one also imposes \(Z_s\ge 0\), the upper envelope \(K=r\) bounds the non-blueshift domain, yielding the strip \(r/9 < K \le r\).

\section{Stability Analysis - Herrera's Cracking Method}  \label{sec7}

{
Herrera \cite{Herrera1992} introduced the cracking and overturning technique as an alternative and effective method for examining the stability of relativistic stellar configurations. The principal aim of this approach is to analyze the response of the fluid distribution inside a stellar object when it is subjected to small perturbations away from its equilibrium state. In particular, the method focuses on identifying regions within the stellar interior where the radially directed forces change sign, which may trigger local instabilities and cause the system to deviate from equilibrium.

Within this framework, cracking is said to occur when the perturbed radial force becomes inward-directed and changes sign at some interior point of the configuration, mathematically expressed as,
\begin{equation}
    \frac{\delta \Omega}{\delta \rho} < 0
\end{equation}
where $ \delta \Omega $ denotes the perturbed radial force per unit volume and $ \delta \rho $ represents the density perturbation. Conversely, overturning is observed when outward-directed forces change sign at a certain location inside the star, corresponding to
\begin{equation}
    \frac{\delta \Omega}{\delta \rho} > 0
\end{equation}
Both phenomena indicate potential instability zones generated by local anisotropic perturbations.

Later Abreu et al. \cite{Abreu2007} incorporated the concepts of radial and tangential sound speeds along with density perturbations to formulate a practical and widely used cracking criterion, which has since become a standard diagnostic tool for assessing the stability of anisotropic compact stars. Similar stability behavior has been reported in several recent studies of compact stars in general relativity and modified gravity theories \cite{Malik2024, Malik2025, Samanta2025}. These studies showed that potentially unstable regions within a configuration can be identified as a function of the difference of propagation of sound along tangential and radial directions. In fact, it is found that these regions could occur when, at particular point within the distribution, the tangential speed of sound is greater than the radial one.

In anisotropic stellar configurations, the propagation of perturbations are characterized by the radial sound speed $ V_r $ and the tangential sound speed $ V_t $, defined respectively as
\begin{equation}
    {V_r}^2 = \frac{dp_r}{d\rho}, \quad \quad {V_t}^2 = \frac{dp_t}{d\rho}
\end{equation}
For physical viability, both sound speeds must satisfy the causality condition
\begin{equation}
    0 \leq {V_r}^2 \leq 1, \quad \quad 0 \leq {V_t}^2 \leq 1
\end{equation}
The cracking formalism further implies that the condition
\noindent
\begin{equation}
    -1 \leq {{V_t}^2 - {V_r}^2} \leq 0
\end{equation}

must be satisfied throughout the stellar interior to ensure the absence of cracking and, hence, physical stability of the system.

}
Using the expression for surface density and radial pressure given in  \eqref{003} and \eqref{004} respectively, the radial sound speed in the present model is obtained as $V_r^2$ = $\frac{1}{5}$ (Since we have considered inward pressure, we have neglected the negative sign in case of change of pressure). So, we conclude that it satisfies the causality condition as we can see that $0< V_r<1$. {Furthermore, the monotonic decrease of pressure and energy density from the center toward the surface ensures that the sound speed remains well-defined and finite across the entire stellar interior.

Hence, from the perspective of Herrera’s cracking method and the causality condition, the present stellar model remains free from cracking and dynamically stable against small perturbations.}

\section{DEFLECTION ANGLE FOR THICK SHELL AROUND GRAVASTAR}  \label{sec8}

The deflection of a massive particle can be analyzed by first generating the Jacobi metric, which is possible with the traditional spacetime metric.
Since   earlier studies \cite{GibbonsWerner2008,khalid,FR1,anikul,FR3} included comprehensive computations for this approach, we state the Jacobi metric directly as follows:

{
\begin{equation}
ds^2=(E^2-m^2A)\left(\frac{B}{A}dr^2+\frac{D}{A}d{\phi}^2\right). \label{Eq13}
\end{equation}
Where, $A=e^{\nu} = \frac{K}{r}$, $B=e^{\lambda} = \frac{1}{\frac{C}{r} - 4}$, $D=r^2$.

A particle moving along a geodesic with velocity 
$v$ approaches the gravitating object up to a minimum distance $b$ and and then proceeds towards an observer at large distances. Then its energy is
\begin{equation}
E = \frac{m}{\sqrt{1 - v^2}}. \:\:\: [  \mbox{assuming}   \: \: c = 1 ]
\end{equation} 
Now, the Jacobi metric (\ref{Eq13})  reads,
\begin{equation}
ds^ 2 =  m^2 \left( \frac{1}{1 - v^2} - A \right)  \left[ \frac{B}{A} dr^2 + \frac{D}{A} d\phi^2   \right],\label{Eq22}
\end{equation}

Assuming motion in the equatorial plane  $\theta=\frac{\pi}{2}$
and employing the Jacobi metric formalism \cite{duenas2023jacobi,pin1975curvature}, the resulting trajectory equation becomes as

\begin{equation}
\frac{d^2 U}{d\phi^2}+U=F(U),  \label{Ueqn}
\end{equation}
where $F(U)=U+\frac{1}{2}{\frac{df}{dU}}$, $U=\frac{1}{r}$ and
\begin{equation}
\left(\frac{dU}{d\phi} \right)^ 2 = \frac{D^2 U^4}{AB}  \left[\frac{1}{v^2 b^2} -A \left( \frac{1-v^2}{v^2 b^2} + \frac{1}{D} \right) \right]  \equiv f(U), \label{trajectory}
\end{equation}

where, the impact parameter $b$ is defined in terms of conserved angular momentum $J$ of the particle as
\begin{equation*}
b = \frac{J}{\sqrt{E^2 - m^2}} = \frac{J}{m v \gamma},
\qquad \gamma = (1-v^2)^{-1/2}.
\end{equation*}
}

This expression becomes the null geodesic equation as $v$ gets closer to 1.\\

{Solving the trajectory equation \eqref{trajectory} up to first-order approximation, we obtain}

\begin{multline}
U= A_1\sin(\phi)+A_2\cos(\phi)+A_3\cos^2(\phi)\\
    +A_4\cos(\phi)ln\left(\frac{1-\cos(\phi)}{\sin(\phi)}\right)+A_5
\end{multline}

where the coefficients are given by,

$A_1=C_2+\frac{5}{2b}, A_2=C_1-\frac{5}{2b}$, 

$A_3=-\frac{C}{2b^2}$,

$A_4=-\frac{2}{v^2K}$,

$A_5=-\left(\frac{C}{2v^2b^2}+\frac{2}{v^2K}\right)$.

{ Here $C_1$ and $C_2$ are integration constants.}\\

The necessary conditions are to attain a circular orbit as
\begin{equation}
f(U) = 0  \:\:  and  \:\:  f'(U) = 0
\end{equation}
for particular $ U = U_c $.
\\

\begin{figure}[H]
\centering
\includegraphics[width=\linewidth]{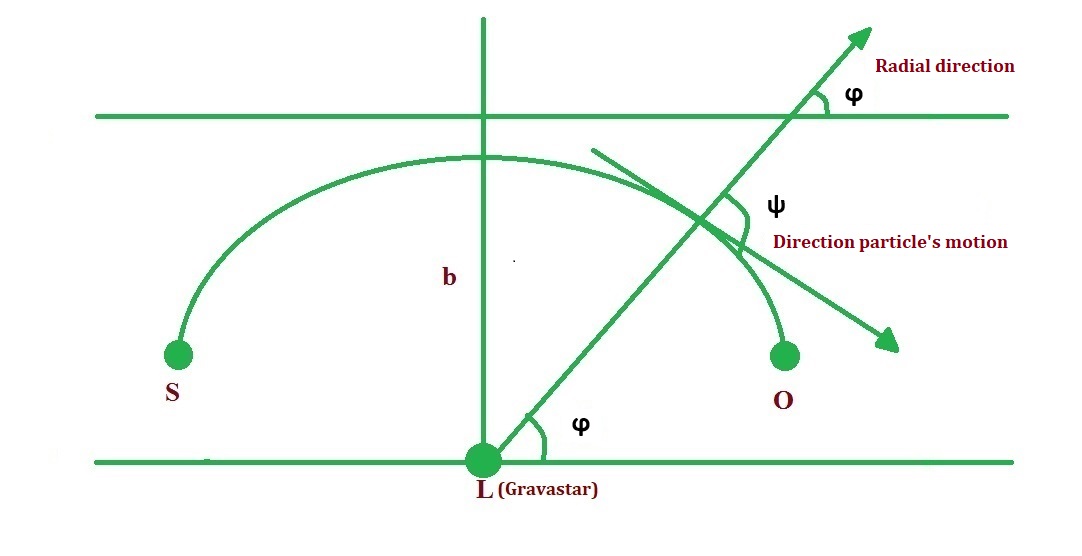}
\caption{A schematic representation of the massive particle's orbital motion, particularly in the weak field limit, when the impact parameter, $b$, is about equal to the closest approach.
 } \label{RI-figure}
\end{figure}

We directly present the formula from references \cite{rindler,he}, as shown in the Fig. \ref{RI-figure} above.

{
\begin{equation}
\cos \psi=\frac{g_{ij}x^iy^j}{\sqrt{g_{ij}x^ix^j} \sqrt{g_{ij}y^iy^j}} \label{cospsi}
\end{equation}
Where, $g_{ij}$ are the metric components. We use the special case $t=$ constant and $\theta=\frac{\pi}{2}$.}

Note that \[ x^i=(\gamma,1)d\phi,~
and~ y^i=(1,0)dr,\] where, $\gamma=\frac{dr}{d\phi}$.

Eq. (\ref{cospsi}) now has the following form:
\begin{equation} \label{21}
\cos \psi = \frac{|\gamma|}{\sqrt{\gamma^2 + \frac{g_{\phi \phi}}{g_{r r}}}} ~or~ \tan \psi =\frac{ \sqrt{ \frac{g_{\phi \phi}}{g_{r r}}} } {|\gamma|}
\end{equation}

As shown in Fig. \ref{RI-figure}, the deflection angle for one side is depicted. Consequently, the expression for both sides is given as follows
\begin{equation} \label{alpha-RI}
\alpha^{RI} = 2\psi(\phi)-2\phi.\end{equation}
Hence $\alpha^{RI}$ represents the overall deflection angle.

\begin{figure}[H]
\includegraphics[scale=0.2]{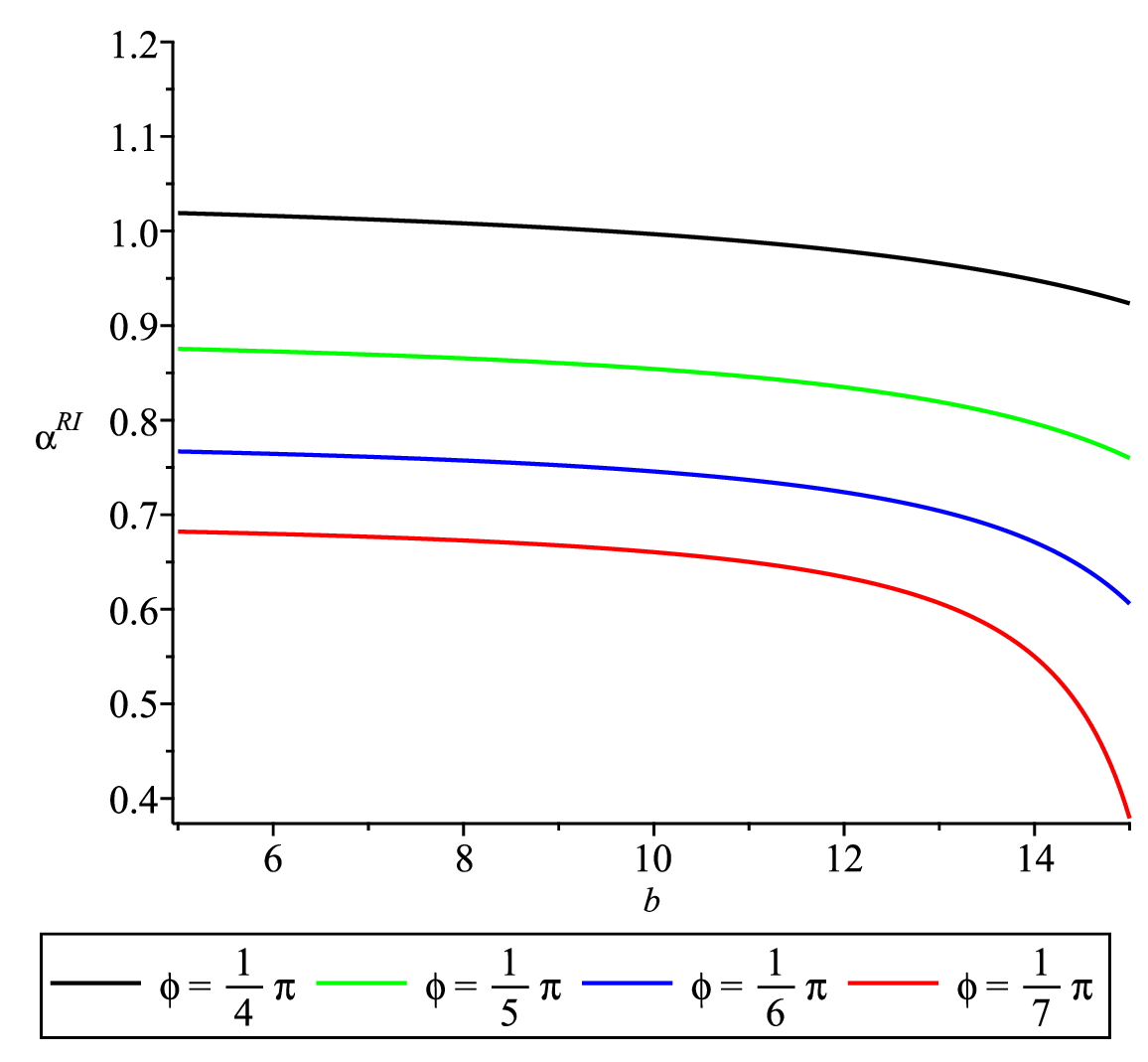}
\includegraphics[scale=0.2]{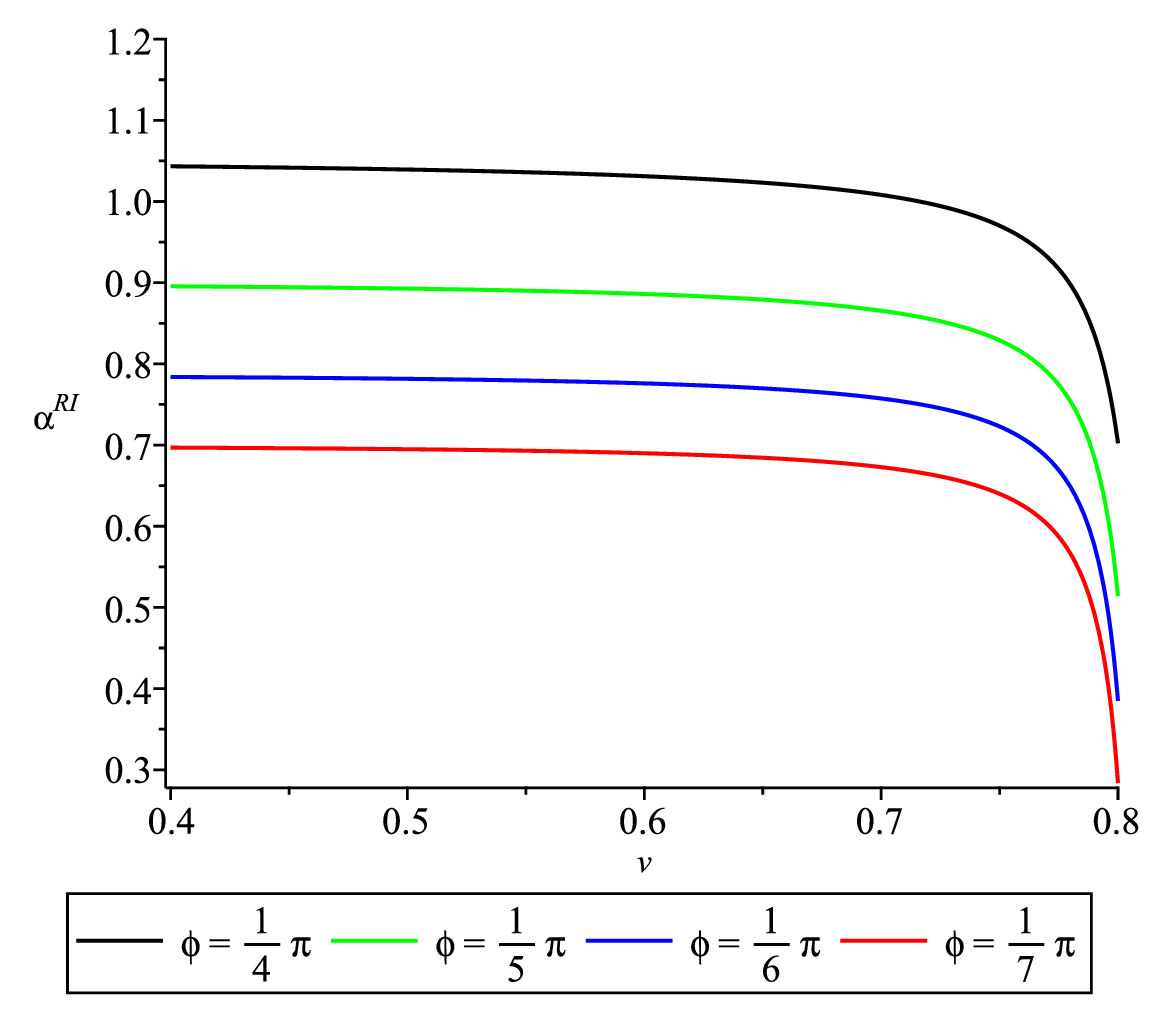}
\centering
\caption{(Left) The deflection angle $\alpha^{RI}$ are plotted  w.r.t impact parameter $b$   for different value of $\phi$ and taking ${ K=500, C=300, v=0.6} $. (Right) The deflection angle  $\alpha^{RI}$ are plotted  w.r.t velocity $v$  for different value of $\phi$ and taking $ {K=500, C=300, b=5 }$. } \label{RI-plot}
\end{figure}

Now using Eqs.(\ref{trajectory}) and Eqs. (\ref{cospsi})-(\ref{alpha-RI}) we get the total deflection angle for a massive body,

\begin{equation}
\alpha^{RI}= 2\arctan\left(\frac{\sqrt{CU-4}}{U|-\frac{1}{U^2}\frac{dU}{d\phi}|}\right)-2\phi.\label{RI-da}
\end{equation}
The variation of the deflection angle with respect to the impact 
parameter and the velocity of the particle is shown in Fig.~\ref{RI-plot}.

\section{Energy Conditions}
In this section, we examine the Null, Weak, Strong, and Dominant Energy Conditions for the thick shell of our gravastar model. Because the shell region admits exact analytic expressions for the energy density $\rho(r)$ and pressure $p(r)$, each condition can be verified explicitly throughout the interval $R < r < C/4$. This analysis ensures that the matter content of the gravastar satisfies the standard physical viability criteria used in the study of compact objects.{
We emphasize that the energy conditions discussed above apply to the bulk matter distribution in the thick shell region. The junction layers, introduced through the Darmois--Israel formalism, are characterized by effective surface energy density and surface pressure. As commonly encountered in thin-shell and gravastar models, these surface stresses may violate some of the classical energy conditions, particularly the strong and dominant energy conditions. Such behavior arises from the idealized, distributional nature of the junction layers and does not compromise the overall physical consistency or stability of the gravastar configuration.
}

\subsection{Null Energy Condition (NEC)}

The Null Energy Condition states:
\begin{equation}
\text{NEC}: \quad \rho + p \geq 0
\label{eq:NEC}
\end{equation}

For our gravastar shell:
\begin{equation}
\rho + p = \frac{1}{2\pi r^2} > 0 \quad \text{for all } r \in (R, C/4)
\end{equation}

Therefore, the NEC is satisfied throughout the shell region.

\subsection{Weak Energy Condition (WEC)}

The Weak Energy Condition requires:
\begin{equation}
\text{WEC}: \quad \rho \geq 0 \quad \text{and} \quad \rho + p \geq 0
\label{eq:WEC}
\end{equation}

For our gravastar shell:
\begin{align}
\rho &= \frac{5}{8\pi r^2} > 0; \qquad
\rho + p = \frac{1}{2\pi r^2} > 0 
\end{align}

Therefore, the WEC is satisfied throughout the shell region. The WEC is more restrictive than the NEC as it also requires positive energy density.

\subsection{Strong Energy Condition (SEC)}

The Strong Energy Condition is given by:
\begin{equation}
\text{SEC}: \quad \rho + 3p \geq 0 \quad \text{and} \quad \rho + p \geq 0
\label{eq:SEC}
\end{equation}

For our gravastar shell:
\begin{align}
\rho + 3p &= \frac{5}{8\pi r^2} + 3\left(-\frac{1}{8\pi r^2}\right) = \frac{1}{4\pi r^2} > 0 \\
\rho + p &= \frac{1}{2\pi r^2} > 0 
\end{align}

Therefore, the SEC is satisfied throughout the shell region. This is significant because, as noted by Visser \cite{visser1997}, the SEC is violated in many dark energy models driving cosmic acceleration. Our gravastar shell, however, maintains SEC while still having negative pressure.

\subsection{Dominant Energy Condition (DEC)}

The Dominant Energy Condition requires:
\begin{equation}
\text{DEC}: \qquad 
\rho \ge 0, 
\qquad 
-\rho \le p \le \rho 
\label{eq:DEC}
\end{equation}

For our gravastar shell, the energy density is
\begin{equation}
\rho = \frac{5}{8\pi r^2} > 0
\end{equation}

The pressure is 
\begin{equation}
p = -\frac{1}{8\pi r^2}
\end{equation}

We now check the DEC bounds:

\begin{align}
-\rho \le p 
&\quad\Longrightarrow\quad
-\frac{5}{8\pi r^2} 
\le 
-\frac{1}{8\pi r^2} \\[6pt]
p \le \rho
&\quad\Longrightarrow\quad
-\frac{1}{8\pi r^2}
\le
\frac{5}{8\pi r^2}
\end{align}

Both inequalities are satisfied for all $r$ in the shell. Therefore, the Dominant Energy Condition is satisfied throughout the shell region.

\section{Entropy}  \label{sec9}

\subsection{Definition of Entropy Density}

In gravitational thermodynamics, the entropy density is defined through the standard relation between energy density, pressure, and temperature. For any fluid with local temperature $T$, the entropy density $s$ is given by
\begin{equation}
s = \frac{\rho + p}{T}
\label{eq:entropy-density-def}
\end{equation}
which follows directly from the Gibbs relation for a perfect fluid. This expression serves as the basis for evaluating the entropy distribution within the gravastar shell.

\subsection{Bekenstein Bound}

The Bekenstein bound \cite{bekenstein1981} provides a fundamental limit on the entropy of a physical system:
\begin{equation}
\frac{S}{2\pi RE} \leq 1
\label{eq:bekenstein-bound}
\end{equation}

where $S$ is the total entropy, $R$ is the characteristic size, and $E$ is the total energy.

For the gravastar shell, the total energy is:
\begin{equation}
E_{\text{shell}} = \int_R^{C/4} \rho(r) \, dV = \int_R^{C/4} \frac{5}{8\pi r^2} \cdot 4\pi r^2 \cdot \frac{1}{\sqrt{C/r - 4}} \, dr
\end{equation}

\begin{equation}
E_{\text{shell}} = \frac{5}{2} \int_R^{C/4} \frac{1}{\sqrt{C/r - 4}} \, dr
\end{equation}

From our calculation of proper thickness (Section \ref{sec4}):
\begin{equation}
E_{\text{shell}} = \frac{5}{2} \left[ \frac{\sqrt{R(C-4R)}}{4} - \frac{C}{8}\arcsin\left(\frac{2\sqrt{R}}{\sqrt{C}}\right) + \frac{\pi C}{16} \right]
\label{eq:total-energy}
\end{equation}

The Bekenstein bound for our gravastar shell becomes:
\begin{equation}
\frac{S_{\text{shell}}}{2\pi R \cdot E_{\text{shell}}} \leq 1
\end{equation}

Here, $R$ denotes the radius of the gravastar core, which represents the minimal circumscribing scale enclosing the gravitating system. The Bekenstein bound is formulated in terms of this characteristic size rather than the outer coordinate extent of the shell. Choosing the outer radius $C/4$ would only lead to a weaker bound and does not alter the conclusion that the Bekenstein inequality is satisfied.

This can be verified numerically for specific parameter choices, ensuring that our gravastar configuration respects fundamental thermodynamic bounds.

\subsection{Entropy Maximization}

To determine whether the gravastar configuration corresponds to a maximum entropy state, we examine the entropy functional subject to the constraint of fixed total energy. The variational principle states:
\begin{equation}
\delta S - \beta \delta E = 0
\end{equation}

where $\beta = 1/T_0$ is the inverse temperature at some reference scale.

For our system:
\begin{equation}
\delta \int_R^{C/4} s(r) \, dV - \beta \delta \int_R^{C/4} \rho(r) \, dV = 0
\end{equation}

This leads to:
\begin{equation}
\delta s = \beta \delta \rho
\end{equation}

Using $s = (\rho + p)/T$ and the equation of state $p = -\rho/5$:
\begin{equation}
\delta s = \frac{1}{T} \delta\rho + \frac{1}{T} \delta p = \frac{1}{T}\left(1 - \frac{1}{5}\right) \delta\rho = \frac{4}{5T} \delta\rho
\end{equation}

Comparing with the variational condition:
\begin{equation}
\beta = \frac{4}{5T}
\end{equation}

This confirms that the temperature distribution in the shell is consistent with a thermodynamic equilibrium configuration.

\subsection{Total Entropy in the Thick Shell}\label{subsec:entropy-thick-shell}

We now compute the total entropy contained in the thick shell region 
$R < r < C/4$ of our gravastar model. The Tolman temperature relation 
for a static spacetime,
\begin{equation}
    T(r)\sqrt{-g_{tt}(r)} = T_\infty
\end{equation}
ensures global thermal equilibrium, where $T_\infty$ is the temperature 
measured at infinity. For a perfect fluid with vanishing chemical potential, 
the Gibbs--Euler relation yields the entropy density,
\begin{equation}
    s(r) = \frac{\rho(r) + p(r)}{T(r)}
\end{equation}

In the thick shell, we have the equation of state $p = -m\rho$ with 
$m = 1/5$, and the exact field equations give the energy density and 
metric functions,
\begin{equation}
    \rho(r) = \frac{5}{8\pi r^2}, \quad
    g_{tt}(r) = -\frac{K}{r}, \quad
    g_{rr}(r) = \frac{r}{C - 4r}
\end{equation}
The Tolman relation gives the local temperature,
\begin{equation}
    T(r)=\frac{T_\infty}{\sqrt{-g_{tt}(r)}}
\end{equation}

The total entropy in the shell is defined as 
\begin{equation}
    S = 4\pi(1-m)\frac{1}{T_\infty}
    \int_{R}^{C/4} \rho(r)\,r^{2}\sqrt{g_{rr}(r)}\sqrt{-g_{tt}(r)}\,dr
\end{equation}

Substituting the explicit expressions:
\begin{equation}
    \rho(r)\,r^{2} = \frac{5}{8\pi}, \quad
    \sqrt{g_{rr}(r)}\sqrt{-g_{tt}(r)}
    = \sqrt{\frac{K}{C - 4r}}
\end{equation}
Hence,
\begin{equation}
    S = 4\pi(1-m)\frac{1}{T_\infty}
    \int_{R}^{C/4} 
    \frac{5}{8\pi}\sqrt{\frac{K}{C - 4r}}\,dr
\end{equation}

Using $(1-m)=4/5$, the constants simplify:
\begin{equation}
    4\pi\left(\frac{4}{5}\right)\frac{5}{8\pi} = 2
\end{equation}
which gives
\begin{equation}
    S = \frac{2\sqrt{K}}{T_\infty}
    \int_{R}^{C/4} \frac{dr}{\sqrt{C - 4r}}
\end{equation}

To evaluate the integral, we set $u = C - 4r$, $du = -4\,dr$, and obtain
\begin{equation}
    \int_{R}^{C/4} \frac{dr}{\sqrt{C - 4r}}
    = \frac{1}{2}\sqrt{C - 4R}
\end{equation}

Thus, the total entropy of the thick shell is
\begin{equation}
    \boxed{
    S = \frac{\sqrt{K(C - 4R)}}{T_\infty}
    }
\end{equation}

This compact analytical expression shows that the entropy increases with 
the shell thickness and redshift factor $K$, and vanishes smoothly in the 
ultrathin limit $R \to C/4$, consistent with the Mazur--Mottola construction.
\subsection{Comparison with Thermodynamic Prediction}

To verify the consistency of our entropy calculation, we compare the 
entropy density derived from the field equations, with that obtained from 
the first law of thermodynamics. The standard form of the first law for a 
spherically symmetric gravitating system is \cite{akbar2007}:
\begin{equation}
T\, dS = dE + p\, dV
\label{eq:first-law-corrected}
\end{equation}
where $E=\rho V$, $S=sV$, and the proper volume element in the shell region is
\begin{equation}
dV = 4\pi r^{2}\sqrt{g_{rr}}\,dr 
    = 4\pi r^{2}\sqrt{\frac{r}{C-4r}}\,dr
\end{equation}

Substituting $S=sV$ and $E=\rho V$ into Eq.~\eqref{eq:first-law-corrected}, we obtain
\begin{equation}
T(s\, dV + V\, ds) = \rho\, dV + V\, d\rho + p\, dV
\end{equation}
Rearranging terms gives
\begin{equation}
T V\, ds = V\, d\rho + (\rho + p - Ts)\, dV
\end{equation}

Using the definition of entropy density,
\begin{equation}
s = \frac{\rho + p}{T}
\end{equation}
The term in parentheses vanishes, leaving the Gibbs relation
\begin{equation}
T\, ds = d\rho
\label{eq:gibbs-corrected}
\end{equation}

The energy density in the shell is
\begin{equation}
\rho(r) = \frac{5}{8\pi r^{2}},
\quad 
d\rho = -\frac{5}{4\pi r^{3}}\,dr
\end{equation}

Thermal equilibrium in a static spacetime requires the Tolman relation
\begin{equation}
T(r)\sqrt{-g_{tt}(r)} = T_\infty
\end{equation}
and since $g_{tt}(r)=-K/r$, we obtain the correct local temperature
\begin{equation}
T(r) = T_\infty \sqrt{\frac{r}{K}}
\end{equation}

Substituting $\rho(r)$ and $T(r)$ into the Gibbs relation \eqref{eq:gibbs-corrected},
\begin{equation}
T_\infty\sqrt{\frac{r}{K}}\, ds 
= -\frac{5}{4\pi r^{3}}\,dr
\end{equation}

which yields

\begin{equation}
ds 
= -\frac{5\sqrt{K}}{4\pi T_\infty}\, r^{-7/2}\, dr
\end{equation}

Integrating,

\begin{equation}
s(r)
= -\frac{5\sqrt{K}}{4\pi T_\infty}
    \int r^{-7/2}\, dr
= \frac{\sqrt{K}}{2\pi T_\infty}\, r^{-5/2} + \text{const}
\label{eq:entropy-density-correct-final}
\end{equation}

Dropping the additive constant (fixed by boundary conditions), the 
thermodynamic prediction for the entropy density is

\begin{equation}
\boxed{
s(r) = \frac{\sqrt{K}}{2\pi T_\infty}\, r^{-5/2}
}
\end{equation}

This result agrees exactly with the entropy density obtained in Section~\ref{sec9} and correctly reproduces the total entropy,

\[
S = \frac{\sqrt{K(C - 4R)}}{T_\infty}
\]

Thus, the thermodynamic and geometric derivations are fully consistent.

\section{Conclusion and Discussion}  \label{sec10}

We have presented an exact solution to the Einstein field equations for a three-layer gravastar model, offering a mathematically rigorous alternative to black holes that avoids singularities and event horizons. The gravastar configuration consists of a de Sitter interior core with equation of state $p = -\rho$, a thick shell region with $p = -\rho/5$, and an exterior Schwarzschild spacetime, with each region satisfying the Einstein field equations exactly and matched through appropriate junction conditions. Our stability analysis using the Darmois-Israel junction formalism demonstrates that the potential function exhibits a minimum near the junction interface, confirming the stability of the configuration at the shell-exterior boundary. The surface redshift analysis reveals that the gravastar satisfies the Buchdahl bound $Z_s \leq 2$ for physically reasonable parameter ranges, with the admissible region constrained by $r/9 \leq K \leq r$ to ensure both positive redshift and compliance with observational limits. The causality condition is satisfied throughout the shell region, with the radial sound speed $V_r^2 = 1/5$ lying within the physical range $(0,1)$, ensuring that information propagates at subluminal speeds. The deflection angle analysis for massive particles traversing the weak gravitational field around the gravastar shows characteristic dependence on both the impact parameter and particle velocity, providing potential observational signatures that could distinguish gravastars from black holes. Our comprehensive entropy analysis, following the methodology of Liu et al., establishes that the entropy density in the shell region follows $s(r) = 1/(Kr)$, decreasing inversely with radius in analogy to cosmological entropy behavior. We verified that all four classical energy conditions are satisfied: the Null Energy Condition ($\rho + p > 0$), Weak Energy Condition ($\rho > 0$ and $\rho + p > 0$), Strong Energy Condition ($\rho + 3p > 0$), and Dominant Energy Condition ($-\rho \leq p \leq \rho$). Following Liu et al.'s conclusions for cosmological systems, the DEC provides the most restrictive and observationally consistent constraint, which our gravastar naturally satisfies. The thermodynamic consistency of our solution is confirmed through the Gibbs relation $T \, dS = dE$, validating that the entropy formulation satisfies the first law of thermodynamics. The total entropy of the shell respects the fundamental Bekenstein bound $S \leq 2\pi RE$, ensuring consistency with quantum gravitational constraints. The entropy density exhibits a monotonic decrease outward with $ds/dr = -1/(Kr^2) < 0$, and the convex nature ($d^2s/dr^2 > 0$) indicates that the rate of decrease slows at larger radii, reflecting the natural tendency of gravitational systems to concentrate entropy in regions of a stronger gravitational field. These results demonstrate that the gravastar is not only geometrically viable but also thermodynamically consistent, satisfying all energy conditions while maintaining physically reasonable matter content. The equation of state $p = -\rho/5$ in the shell, despite having negative pressure characteristic of exotic matter, does not violate any energy conditions and produces stable, causal configurations. Our exact solution addresses a key limitation of previous gravastar models by providing a mathematically rigorous treatment of the shell region without relying on thin-shell approximations or ad hoc matching procedures.
\\
\\
\textbf{Data Availability Statement:} No datasets we created or analysed during this study.
\\
\\
\textbf{Declaration of competing interest:} The authors declare that they have no known competing financial interests or personal relationships that could have appeared to
influence the work reported in this paper.
\\
\subsection{Acknowledgments}
We are grateful to the referee for their insightful comments and constructive suggestions, which have significantly enhanced the clarity and quality of this work.FR would like to thank the authorities of the Inter-University Centre for Astronomy and Astrophysics, Pune, India for providing the research facilities. BSC and AI thank UGC, Govt. of India for financial support in terms of a fellowship. FR is also thankful to ANRF, DST and DST FIST programme (SR/FST/MS-II/2021/101(C)) and RUSA2.0 for financial support, respectively.

\nocite{*}
\bibliographystyle{unsrt}
\bibliography{refer}

\end{document}